# Bimodal Plasmonic Refractive Index Sensors Based on SU-8 Waveguides

Omkar Bhalerao[1,2], Stephan Suckow[1*], Horst Windgassen[1], Harry Biller[3], Konstantinos Fotiadis[4], Stelios Simos[4], Evangelia Chatzianagnostou[4], Dimosthenis Spasopoulos[4], Pratyusha Das[5], Laurent Markey[5], Jean-Claude Weeber[5], Nikos Pleros[4], Matthias Schirmer[3], Max C. Lemme[1,2,*]

[1]AMO GmbH, Otto-Blumenthal Str. 25, Aachen, 52074, Germany

[2]Chair of Electronic Devices, RWTH Aachen University, Otto-Blumenthal Str. 25, Aachen, 52074, Germany

[3]Allresist GmbH, Am Biotop 14, Strausberg, 15344, Germany

[4]CIRI-AUTH, Balkan Center, Building A, 10th km Thessaloniki-Thermi Rd., Thessaloniki, 57001, Greece

[5]Laboratoire Interdisciplinaire Carnot de Bourgogne (ICB), CNRS UMR 6303, Université de Bourgogne, BP 47870, Dijon, 21078, France.

* Email: suckow@amo.de, max.lemme@eld.rwth-aachen.de

**Abstract**

Plasmonic refractive index sensors are essential for detecting subtle variations in the ambient environment through surface plasmon interactions. Current efforts utilizing CMOS-compatible, plasmo-photonic Mach-Zehnder interferometers with active power balancing exhibit high sensitivities at the cost of fabrication and measurement complexity. Alternatively, passive bimodal plasmonic interferometers based on SU-8 waveguides present a cost-effective solution with a smaller device footprint, though they currently lack opto-mechanical isolation due to exposed photonic waveguides. In this work, we introduce innovative polymer-core and polymer-cladded bimodal plasmonic refractive index sensors with high refractive index contrast. Our sensors feature an aluminum stripe, a bilayer SU-8 photonic waveguide core, and the experimental optical cladding polymer SX AR LWL 2.0. They achieve a sensitivity of (6300 ± 460) nm/RIU (refractive index unit), surpassing both traditional and polymer-based plasmo-photonic sensors. This approach enables integrated, wafer-scale, CMOS-compatible, and low-cost sensors and facilitates plasmonic refractive index sensing platforms for various applications.

1. **Introduction**

In recent years, the field of plasmonics has experienced remarkable growth, finding applications in biomedical sensing, chemical analysis, food processing, and quality control [1], [2], [3], [4]. Although traditional photonic waveguide (WG) refractive index (RI) sensors, such as fiber optic sensors, offer high reliability, they are limited in sensitivity and cannot be miniaturized [5], [6]. Plasmonic RI sensors provide sensitive operation by exposing the electric field to the surrounding medium, making them suitable for miniaturization into photonic integrated circuits (PICs). This allows for leveraging their unique properties for enhanced functionality, thereby enabling versatile applications. The highly sensitive active plasmonic area in a PIC can further be functionalized to interact with a broad array of analytes, enabling real-time, label-free, and spectroscopic detection [7]. Such plasmo-photonic devices, primarily based on gold (Au) and silver (Ag), have initially been used for RI sensing on silicon-on-insulator (SOI) [8], [9], [10] and silicon nitride (SiN) photonic WG platforms [11], [12], combining high plasmonic sensitivity with low-loss WGs. However, substituting Au with aluminum (Al) as the plasmonic material facilitates plasmonics integration within silicon (Si) CMOS technology [13]. Such CMOS compatibility, in combination with its relative abundance and broadband response, make Al an appealing choice for various plasmonic sensing applications [14], [15]. Hence, we have recently demonstrated plasmo-photonic Mach-Zehnder interferometers (MZIs) on the SiN WG platform with integrated Al stripes in the sensing arm, achieving high sensitivities of up to 4760 nm/RIU [16]. These sensors utilize SiN WGs cladded with $SiO_2$, resulting in low optical losses and CMOS compatibility. However, the multi-step fabrication processes required for SiN WGs increase both fabrication complexity and costs, posing specific challenges for disposable sensing applications.

An alternate approach is to use all-polymeric WGs. The cost-efficient polymer SU-8 offers excellent chemical stability, good optical transparency above 400 nm wavelength, and simple, one-step lithography for WG definition [17], [18], [19]. Bimodal Al plasmo-photonic interferometers based on SU-8 WGs have demonstrated competitive sensitivities of 4460 nm/RIU [20], eliminating the requirement for active power balancing and offering a smaller footprint. However, they currently suffer from severe susceptibility to the external environment due to the lack of a cladding layer over SU-8 WG core. This reduces the sensors' reliability and limits their compatibility with microfluidics, a typical integrated photonic sensor technology. Hence, cladding layers are crucial for opto-mechanical isolation and reliable device operation.

Typical polymer claddings result in refractive index (RI) contrasts of 0.05 and 0.01 to the SU-8 WG at the operational wavelength of 1.55 µm. This low contrast leads to weak light confinement within the SU-8 core, which increases the risk of parasitic slab modes in the cladding layer and limits the photonic device options. Polymethyl methacrylate (PMMA) offers significantly higher RI contrast. However, PMMA is typically exposed through deep ultraviolet (DUV; λ = 200 nm – 270 nm) lithography. Extending lithography to the i-line regime (λ = 365 nm) for cheap exposures requires the addition of a suitable chemical agent [21]. Alternative "glass-like" polymers like Spin-on Glass (SOG) also provide higher RI contrast to the SU-8 core. However, standard SOG requires drying and curing steps to form sufficiently thick homogenous layers, which are usually performed at temperatures around 350°C. This renders them incompatible with SU-8 processing. Lastly, traditional cladding materials like optical-grade fused silica glass ($SiO_2$) yield RI contrasts of approximately 0.13, enabling the fabrication of components similar to conventional PICs. Standard deposition process of $SiO_2$ layers typically involve temperatures above 200°C, oxygen ($O_2$) plasma, or both, which adversely affect SU-8 due to its susceptibility to $O_2$ plasma and high thermal budget, severely limiting direct deposition of a sufficiently thick $SiO_2$ cladding layer on SU-8 WG core. Additionally, many applications require a patterned cladding layer with openings to access electrical contact pads, create vias, or reveal sensing areas. When using reactive ion etching (RIE) of $SiO_2$, only specific fluorine-based chemistries providing limited selectivity can be used with the SU-8 core [22]. Wet chemical etching of $SiO_2$ is often done with hydrogen fluoride (HF), which is corrosive towards SU-8 and, therefore, not suitable for applications requiring precise feature size control. Moreover, HF tends to creep into interfaces, leading to undesired delamination of other layers.

Here, we introduce an experimental cladding material, SX AR LWL 2.0 (Allresist GmbH), that is compatible with SU-8 processing. This material enabled the creation of polymer-core/polymer-cladded bimodal plasmonic RI sensors that can be readily integrated within PICs and microfluidic systems. The sensors were optically characterized at a wavelength of 1.55 µm, confirming the interferometric spectrum and length-dependent scaling of the free spectral range (FSR) in ambient air. The sensors' sensitivity was obtained by exposing them to liquids of varying RI. Simulations support the experimental FSR and sensitivity values, which stand out amongst other bimodal plasmo-photonic RI sensors. The targeted device application as an ultra-sensitive biosensor designed to detect contaminations in the vegetable supply chain [23], was experimentally simulated using a range of analytes with differing RI.

## 2. Sensor concept and simulations

*Concept*

The bimodal plasmonic interferometers consist of an SU-8 WG featuring an Al stripe at its center (Figure 1). This design required the integration of the plasmonic Al layer within the SU-8 WG. We achieved this by consecutively depositing two layers of SU-8 (for details, see fabrication section). The SU-8 WGs have a total thickness of 1.8 µm and a width of 7 µm, optimized through ANSYS Lumerical [24] simulations for exciting only the fundamental modes of the WG, at least vertically ([20], [24], Figure 1a). They are covered by a ca. 6 µm thick SX AR LWL 2.0 cladding layer, opened solely at the sensing areas bearing the plasmonic stripe. The fundamental TM photonic mode is excited in the SU-8 WG and converted into dual plasmonic modes on both the top and bottom sides of the Al stripe (see Figure 1b). On the upper side, the Al/air interface facilitates the distinct propagation of a "top" plasmonic mode. Simultaneously, the Al forms an interface with the remaining lower SU-8 layer, generating a separate "bottom" plasmonic mode. The exposure to the ambient atmosphere and the corresponding change in RI introduces a phase shift in the top mode, while the bottom mode remains unaffected, establishing the configuration of a bimodal plasmonic interferometer. The top mode then functions as the sensing arm, where changes in the surrounding medium induce further phase shifts. The two plasmonic modes propagate independently along the surfaces of the Al stripe and interfere solely upon coupling back into the SU-8 WG. This bimodal interferometric transmission spectrum traverses through the bilayer SU-8 WG stack to the output facet.

*Simulations*

Simulations were performed with ANSYS Lumerical to analyze the power coupling between the photonic and the plasmonic part (MODE), plasmonic propagation, and overall device operation (INTERCONNECT). The simulations focused on maximizing sensitivity and optimizing the extinction ratio (ER) at the sensor output. The optimization strategy involved adjusting coupling coefficients from the photonic to the two plasmonic modes and tuning the plasmonic stripe length to achieve optimal interference and, thus, ER. A thickness variation of 0.5 µm to 1.3 µm of the bottom SU-8 layer served to control coupling coefficients, impacting the overlap integral between the TM photonic mode and the two plasmonic modes. The 2D eigenmode analysis considered the overlap integrals and propagation losses of the top and bottom plasmonic modes, dependent on the bottom SU-8 layer thickness. Details of the simulations can be found in the supporting information and in our previous work [20].

3. Fabrication

All devices were fabricated on 150 mm Si substrates containing a 3 µm thick thermally grown $SiO_2$ layer. Our design foresees the plasmonic Al layer to be embedded in the SU-8 WGs. Hence, we chose a bilayer SU-8 process [25]. Initially, the two layers were developed at the same time by an appropriate developer. However, the deposition and patterning of the Al plasmonic stripes on the exposed but undeveloped bottom SU-8 layer led to excessive wrinkling of the SU-8, similar to [26]. We thus changed the process to individually fabricated bilayer SU-8 WGs. This approach allowed separate processing and hard-baking of the SU-8 layers. The hard-bake processes ensured the mechanical and chemical stability of the resulting layers, facilitating the integration of various fabrication steps between the photonic WG formation steps. Additionally, processing each layer separately significantly reduced the risk of Al contamination on the photonic-to-plasmonic coupling facets and the SU-8 sidewalls.

The bi-layer WG stack was formed using the commercially available SU-8 2 formulation (micro resist technology GmbH). To form the bottom WG layer, 6 mL of SU-8 2 were deposited on the wafer using a manual spin-coater, resulting in a 900 nm thick film. A two-stage soft bake at 65°C and 95°C prepared the layer for lithography. After the wafer cooled to room temperature, a Canon FPA 3000 i5+ stepper tool performed i-line projection lithography, followed by a two-stage post-exposure bake (PEB) with parameters identical to the soft bake. The wafer was immersed in propylene glycol monomethyl ether acetate (PGMEA) to dissolve the unexposed

SU-8. After rinsing in isopropyl alcohol (IPA) and deionized (DI) water, a hard bake at 150°C formed the bottom SU-8 WGs. The RI of SU-8 at 1.55 µm was around 1.57 and was determined theoretically by the Cauchy equations provided by the manufacturer (see supporting information Figure S1).

After processing the bottom WG layer, an 80 nm thick Al layer was deposited using an FHR Star 200 EVA electron beam evaporator. The coating and development steps for patterning the plasmonic stripes using AZ 5214E photoresist were optimized to accommodate the underlying topography created by the SU-8 layer. The mask design included sensors of varying plasmonic stripe lengths (25 µm – 200 µm) to verify the length-dependent scaling of FSR (see optical characterization section). Following i-line projection lithography, the resist was developed in a tetramethyl ammonium hydroxide (TMAH)-based developer solution. TMAH developer, with typically 2% - 3% concentration is known to etch Al at the rate of 50 nm - 100 nm per minute [27]. Therefore, the recipe was optimized for minimal consumption of the Al layer during development. The resist pattern thus obtained served as the etch mask for a subsequent Al wet-etch process in a commercially available buffered phosphoric acid solution. After etching, the resist mask was removed by immersing the wafer in acetone followed by a wash in IPA and DI water. The use of wet chemistry maintained the silicon technology compatibility of our sensors and ensured metal-free sidewalls in the bottom SU-8 layer.

The top SU-8 layer was processed like the bottom layer. All process parameters remained identical except for the spin-coating, which was adjusted to accommodate the topography formed by the bottom SU-8 WG and the Al plasmonic stripe. The resulting WG stack was also hard baked at 150°C. The process concluded with a DI water rinse followed by a drying cycle. As the bottom SU-8 experiences twice the exposure dose and processing temperature cycles as the top layer, a RI difference of approximately $1 \times 10^{-3}$ was observed between the SU-8 layers.

The cladding layer processing was performed as per manufacturer specifications, starting with the spin-coating of an adhesion promoter, SX AR 300-80/12 (Allresist GmbH), followed by a baking step performed at 180°C. After cooling down, SX AR LWL 2.0 was spin-coated, followed by a soft baking step at 85°C. Next, an i-line exposure and PEB step, identical to the soft bake, were followed by immersion development in the AR 600-546 (Allresist GmbH) developer. Immersion in IPA facilitated developer stop, rinse, and removal. The manufacturer-provided RI of the cladding layer at 1.55 µm wavelength, measured using a Metricon Model 2010/M prism coupler, was approximately 1.42 (see supporting information Figure S1). The experimental cladding polymer used in this study was not commercially available at the time of this publication but is expected to be offered in due time. An overview of the entire fabrication process is provided in Figure 2. After fabrication, the wafers were diced into individual chips measuring 2 cm x 2 cm. The dicing recipe was optimized to produce smooth coupling facets in the cladded bilayer SU-8 WGs.

### 4. Optical characterization

The measurement setup consisted of a TM-polarized, lensed, polarization-maintaining single-mode fiber (SMF) connected to a tunable laser operating around the 1.55 µm wavelength for launching a fundamental TM mode into the SU-8 WG stack (see Figure 3a). At the output facet, a second SMF connected to a power meter extracted the optical transmission spectrum. Both input and output fibers were butt-coupled to the chip. The laser input power remained constant at 1 mW while the input wavelength was varied between 1.50 µm - 1.63 µm in 0.1 nm steps for all measurements. There was no thermal stabilization applied during optical characterization. Initially, an optical link was established using reference SU-8 WGs without plasmonic features. Since there is no photonic-to-plasmonic conversion in these WGs, they serve to establish proper fiber alignment and reliable photonic operation. Alignment of the input and output fibers was verified by initially launching a visible red laser beam (λ~635 nm) through the input fiber into the reference SU-8 WGs. Confirmation of an optical link was then achieved by switching to the tunable laser source (λ=1.55 µm) and maximizing the transmission through the reference SU-8 WGs. Optical transmission measurements were then conducted in ambient air on the reference WGs and all plasmonic sensor lengths to obtain the reference and plasmo-photonic transmission spectra, respectively. These spectra were used for subsequent analysis of sensor response and bimodal interferometric behavior. The bimodal interferometric spectrum in the sensors arises from the interaction of the top plasmonic mode with air (n=1) and the bottom plasmonic mode with SU-8 (n~1.57), resulting in a phase shift between them due to the differing RI. This phase shift leads to characteristic minima positions in the transmission spectrum, confirming interference between both plasmonic modes and validating the bimodal configuration. All sensors displayed distinct minima in their transmission spectra, indicating bimodal interference, and sensors of identical length exhibited comparable interferometric transmission spectra without requiring data filtering or smoothing (see Figure S2a). Additionally, the median optical transmission decreased with increasing plasmonic stripe lengths, indicating the excitation and propagation of the plasmonic modes (see Figure S2b).

The FSR of the sensors is calculated theoretically by the following equation:

$$FSR = \frac{\lambda^2}{\Delta n_g \cdot L} \tag{1}$$

Where λ is the operation wavelength, $\Delta n_g$ is the group index difference between the top and bottom plasmonic modes, and L is the length of the plasmonic stripe. The optical transmission spectra of all sensor lengths showed FSR variations consistent with simulated values, confirming the bimodal interferometer experimentally (see Figure 3b). Meanwhile, the ER of the sensors is obtained theoretically by the following equation:

$$\mathrm{ER} = 10 \log_{10} \frac{\mathrm{T_{max}}}{\mathrm{T_{min}}} \qquad (2)$$

Where $T_{max}$ and $T_{min}$ are the maximum and minimum values of the interferometer power at the output. ER variations corresponding to changes in the plasmonic sensor length were also observed (see Figure 3c). Sensors up to 100 µm in length exhibited an ER exceeding 10 dB, qualifying them as potential practical sensors. Among these, the 75 µm long sensor demonstrated the highest ER, suggesting an optimal power balance between the plasmonic modes. However, the FSR of the 25 µm and 50 µm sensors lay outside and very close to the edges of our observation window, respectively. Additionally, the low ER and spectral ripples of sensors ranging from 125 µm to 200 µm in length made the identification of FSR unreliable without dedicated algorithms. As a result, 75 µm and 100 µm sensor lengths were chosen for subsequent repeatability and RI sensing analysis. Multiple 75 µm sensors were characterized in air, and the minima positions in their transmission spectra were recorded. The consistency of these minimum positions indicates reliable generation of the bimodal interferometric spectrum. The data aggregated from 28 different 75 µm sensors is presented in Figure 3d. The average deviation from the mean minimum positions in the transmission spectra was within ± 13 nm, and the range between the maximum and minimum values was within ± 21 nm.

After concluding the repeatability analysis, initial RI sensing experiments were conducted on 75 µm sensors. A water droplet was placed on the sensors, and the optical transmission spectra were recorded. The obtained transmission data was normalized to eliminate the influence of photonic WG propagation, thereby revealing the actual sensing response. For normalization, multiple reference SU-8 WGs were initially examined, and the transmission spectra of four such WGs were averaged to generate a reference transmission spectrum (see Figure S3). Subsequently, the sensor insertion losses were calculated by subtracting the reference spectrum from the optical transmission spectrum of each sensor. The sensor insertion losses primarily represent the photonic-to-plasmonic conversion loss at each facet and the plasmonic propagation loss along the sensor length. The transmission spectrum of a 75 µm long sensor in water is shown in Figure 3e. Furthermore, within the same figure, the transmission spectrum of the same sensor alongside an identical-length sensor placed in air is also illustrated. It was observed that the FSR changed significantly after exposing the sensor to water, in agreement with the simulations. The variation in FSR can be attributed to the change in the ambient RI affecting the relative phase difference between the top and bottom plasmonic modes, also altering the interference conditions as the top mode is exposed to water (n=1.311 [28]) instead of air (n=1). Therefore, RI sensing was validated, and subsequently, the 100 µm long sensors were subjected to liquids with varying RI and their transmission spectra were recorded (see Figure 3f). The RI of each liquid was verified using a commercial refractometer prior to performing the on-chip experiments. As the sensor length was maintained at 100 µm for all RI measurements, a blue shift in the transmission minimum was observed with increasing analyte RI. The shift in minima positions was plotted against their corresponding RI values, as shown in

Figure 3g. The experimental bulk sensitivity ($S_{bulk}$) was obtained from the slope ($\Delta\lambda_{minima}/\Delta n$) of the linear fit performed on this plot, while the theoretical bulk sensitivity was calculated from the following equation:

$$S_{bulk} = \frac{\Delta \lambda_{minima}}{\Delta n} \tag{3}$$

Where $\Delta\lambda_{minima}$ indicates the shift in the minimum position induced by the change $\Delta n$ in the ambient RI. The experimental value of $S_{bulk}$ obtained from 100 μm and 75 μm long sensors were (6300 ± 460) nm/RIU and (5390 ± 464) nm/RIU, respectively, both closely matching the simulated value of 6080 nm/RIU [29]. The limit of detection (LoD) is the lowest concentration of the analyte (RIU) that can be reliably measured by a sensor and is calculated as follows:

$$LoD = \frac{R}{Sensitivity} \tag{4}$$

where R is the smallest possible spectral shift that can be measured with the employed experimental setup. The resolution of the tunable laser used for our sensitivity measurements is 0.0001 nm. Considering stability of the measurement setup and system noise, we assume R to be 0.01 nm [30]. The sensitivity is derived from the experiment with varying RI liquids (see Figure 3g). From the experimental data, a LoD of 1.58 x 10$^{-6}$ RIU was estimated for the 100 μm long sensors. While multiple strategies for implementing a plasmo-photonic RI sensor exist [31], our approach targets the exploitation of the bimodal behavior on a metallic plasmonic WG operating at the 1.55 μm telecommunication wavelength. In this regard, our sensor has demonstrated the highest sensitivity (see Table 1) among other bimodal plasmonic RI sensors based on both Au and Al platforms.

Table 1: Performance of different fabricated plasmo-photonic RI sensors.

| Interferometer Type | Plasmonic Material | Bulk Sensitivity (nm/RIU) | LoD (RIU) | CMOS Compatibility | Operation Wavelength (μm) |
|---|---|---|---|---|---|
| Bimodal stripe WG interferometer [32] | Au | 435 | 1 x 10$^{-3}$ | No | 1.55 |
| Bimodal plasmonic interferometer [33] | Au | 2430 | NA | No | 1.55 |
| Bimodal plasmonic interferometer [20] | Al | 4464 | 2.1 x 10$^{-6}$ | Yes | 1.55 |
| **This work** | **Al** | **6300** | **1.5 x 10$^{-6}$** | **Yes** | **1.55** |

## 5. Conclusions

In conclusion, our work successfully introduces and validates bimodal plasmonic RI sensors featuring a polymer-core and polymer-cladding configuration. The plasmonic sensors are created by embedding an Al stripe within a bilayer SU-8 WG stack. The plasmonic stripe is exposed to the ambient from the top side, where it can be functionalized to interact with diverse analytes. The excitation of separate plasmonic modes on both the upper and lower sides of the Al stripe form the bimodal interferometer configuration. The bilayer SU-8 WG stack is cladded with the experimental SX AR LWL 2.0 optical polymer (Allresist GmbH). This novel cladding layer is crucial to the devices' operation, as it provides superior opto-mechanical isolation of the SU-8 core and an RI contrast of approximately 0.14. Optical characterization conducted at the telecommunication wavelength of 1.55 µm demonstrates excellent agreement between simulated and experimental values of ER, FSR and sensitivity. Sensing experiments with liquids of varying RI show remarkable bulk sensitivity of (6300 ± 460) nm/RIU on 100 µm long sensors. From the experimental data, we estimate a LoD of 1.58 x 10$^{-6}$ RIU, confirming the sensor's record-high performance among other bimodal plasmo-photonic RI sensors. Our work demonstrates wafer-scale, CMOS-compatible, extremely sensitive, and cost-effective plasmonic RI sensors, ready for integration with traditional microfluidic-based photonic interrogation platforms.

**Acknowledgements**

We are grateful to Prof. Dr. Anna Lena Schall-Giesecke and Dr. Piotr Cegielski for discussions. We also express our appreciation to Bartos Chmielak, Caroline Porschatis, and Holger Lerch for their expert advice and assistance with device fabrication. This work has received funding from the European Union's Horizon 2020 research and innovation programme under the grant agreement number 101007448 (GRACED). This work has been partially funded by the French Agence Nationale de la Recherche (EIPHI Graduate School ANR-17-EURE-0002) with the support of the French Agence Nationale de la Recherche under program Investment for the Future (ANR-21-ESRE-0040), the Région de Bourgogne Franche-Comté, the European Regional Development Fund (FEDER-FSE Bourgogne Franche-Comté 2021/2027), the CNRS and the French RENATECH+ network.

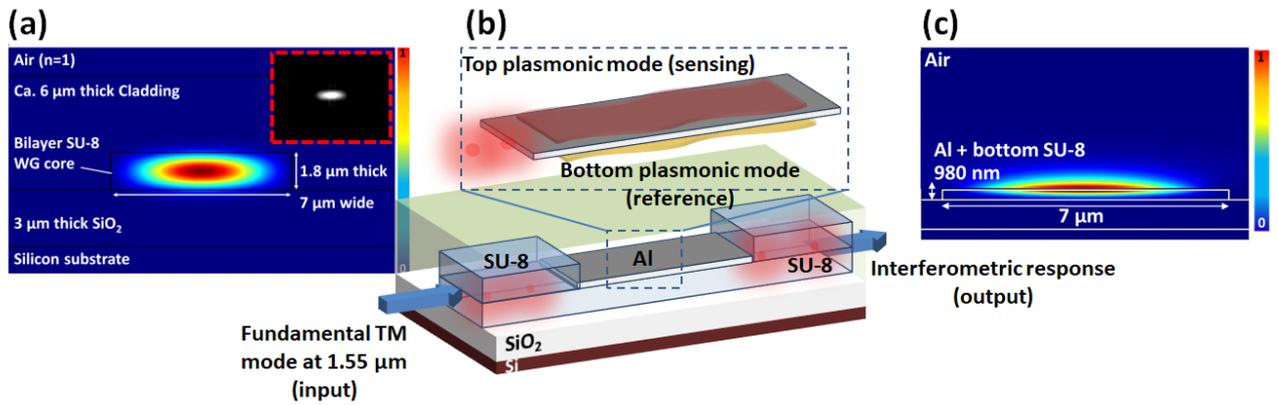

Figure 1: Schematic of the bimodal interferometer concept. (a) Simulated mode field profile of the fundamental TM mode within a 1.8 µm x 7 µm (thickness x width) bi-layer SU-8 WG stack. Butt-coupling was employed for injecting the fundamental TM mode and for recording the optical transmission spectrum. The inset shows the observed mode field profile at the outcoupling facet of a polymer-cladded reference SU-8 WG without plasmonics, indicating single-mode operation. (b) A 3D illustration of the bimodal interferometer. The excitation scheme of the top and bottom plasmonic modes on an Al stripe is shown in an inset. (c) Simulated cross-sectional composite image of the top and bottom plasmonic modes on an Al stripe. The Al stripe with dimensions 80 nm x 7 µm (thickness x width) is positioned on top of a 900 nm x 7 µm (thickness x width) bottom SU-8 layer.

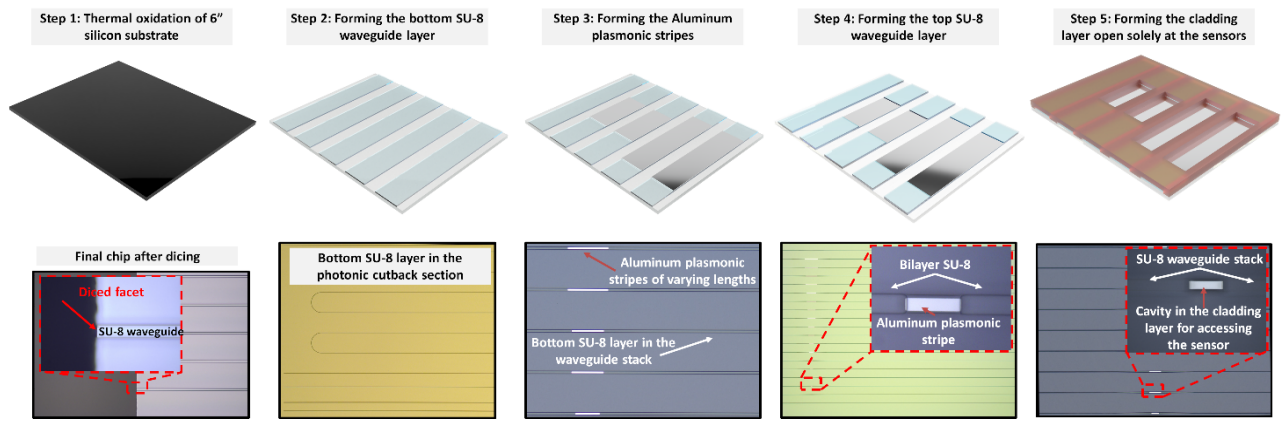

Figure 2: The major fabrication processes of the bimodal interferometer, along with optical microscope images at each step. The entire process flow was developed to ensure simple, wafer-scale, scalable, and CMOS compatible fabrication. The bilayer SU-8 WG stack formation, bimodal Al plasmonic stripe fabrication, along with the final dicing step for generating coupling facets are critical to the functionality of the sensor.

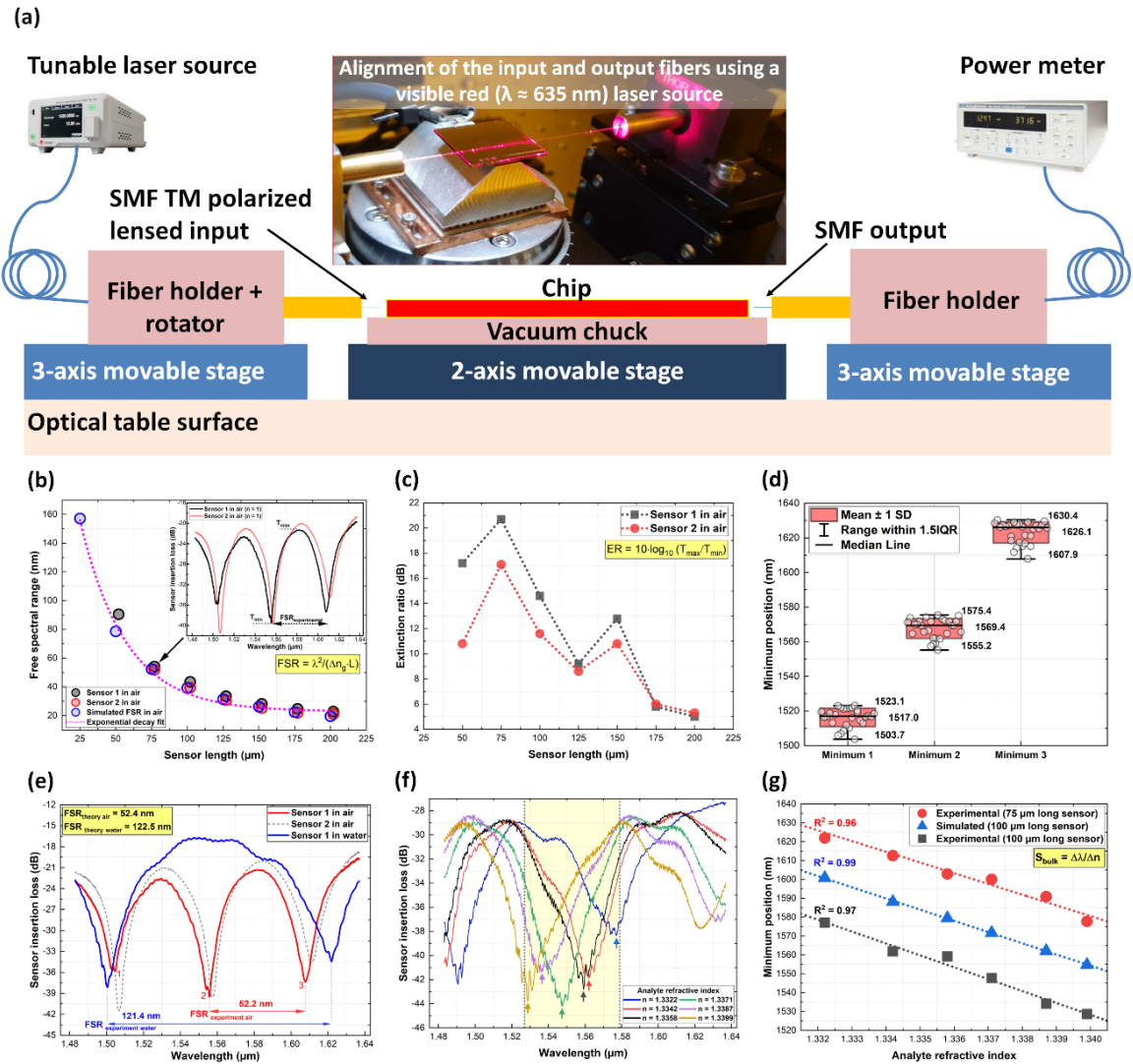

Figure 3: Optical characterization setup and results of the bimodal interferometers. (a) Schematic of the measurement setup used for optical characterization. The optical characterization was performed around the 1.55 µm telecommunications wavelength. (b) The theoretical and experimental FSR values for various sensor lengths obtained from the optical transmission measurements shown in Figure S3a. A clear plasmonic-length dependent scaling was observed, aligning strongly with the simulated values. The exponential decay fit is made only for improved visualization. The inset shows the optical response of two 75 µm long sensors in air, indicating minimum positions in the transmission spectrum and the corresponding transmission values utilized for determining FSR and ER. Both spectra show clear minima, and no smoothing was applied to the curves. (c) The experimental ER recorded

from different sensor lengths. ER for 25 µm long sensors could not be calculated as the maximum and minimum positions in the transmission spectrum lied outside the interrogation window of the setup. (d) A boxplot representing the data obtained from the minimum positions in the optical transmission spectra of 28 different 75 µm sensors showing the repeatability of sensing response in ambient air. (e) The optical transmission spectrum of the 75 µm sensors exposed air and water. The minimum positions used for sensor repeatability analysis in Figure 3d have been numbered for easy identification. (f) The optical transmission spectra of a 100 µm sensor exposed to liquids of varying RI. The highlighted area denotes the spectral range used for analysis, while the arrows denote the minimum positions considered for plotting the sensitivity. (g) The experimental sensitivity calculations of 75 µm and 100 µm sensors. The minimum positions in their transmission spectra were plotted against the corresponding RI values and a linear fit was applied, with the slope representing the sensitivity value. The simulated sensitivity of the 100 µm sensors is also provided for reference.

# Supporting information

# Bimodal Plasmonic Refractive Index Sensors Based on SU-8 Waveguides

Omkar Bhalerao[1,2], Stephan Suckow[1,*], Horst Windgassen[1], Harry Biller[3], Konstantinos Fotiadis[4], Stelios Simos[4], Evangelia Chatzianagnostou[4], Dimosthenis Spasopoulos[4], Pratyusha Das[5], Laurent Markey[5], Jean-Claude Weeber[5], Nikos Pleros[4], Matthias Schirmer[3], Max C. Lemme[1,2,*]

[1]AMO GmbH, Otto-Blumenthal Str. 25, Aachen, 52074, Germany

[2]Chair of Electronic Devices, RWTH Aachen University, Otto-Blumenthal Str. 25, Aachen, 52074, Germany

[3]Allresist GmbH, Am Biotop 14, Strausberg, 15344, Germany

[4]CIRI-AUTH, Balkan Center, Building A, 10th km Thessaloniki-Thermi Rd., Thessaloniki, 57001, Greece

[5]Laboratoire Interdisciplinaire Carnot de Bourgogne (ICB), CNRS UMR 6303, Université de Bourgogne, BP 47870, Dijon, 21078, France.

* Email: suckow@amo.de, max.lemme@eld.rwth-aachen.de

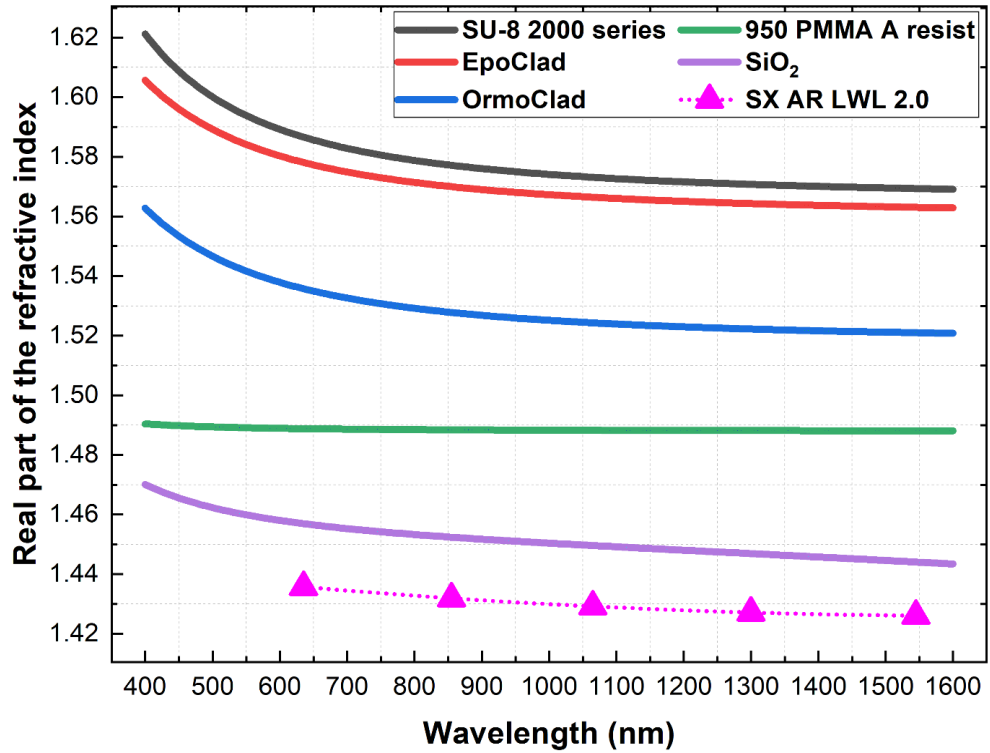

S1: RI values of various photonic materials reported in this publications. The real part of the RI of SU-8, EpoClad, OrmoClad, PMMA, $SiO_2$ (fused silica), and the experimental SX AR LWL 2.0 cladding at various wavelengths. The RI values for SU-8 [34], EpoClad [35], OrmoClad [36], and PMMA [37] were derived theoretically using the Cauchy equations provided by their respective manufacturers. The RI values for $SiO_2$ were obtained from the dispersion relation as reported in literature [38]. The experimental RI values for the SX AR LWL 2.0 cladding were measured using a Metricon Model 2010/M prism coupler and were supplied by the manufacturer.

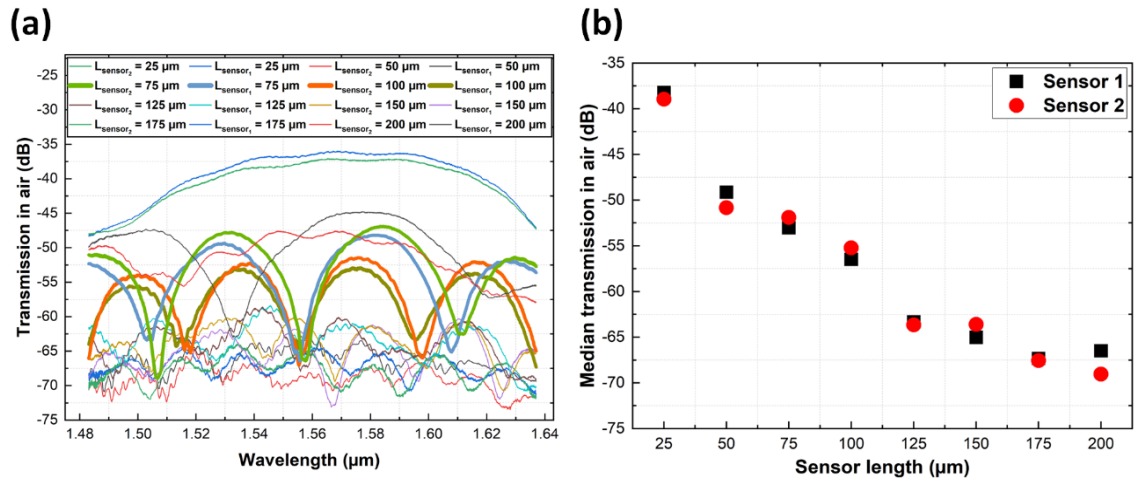

S2: Optical transmission measurements of the bimodal interferometers in air. (a) The optical transmission spectra belonging to sensors of varying lengths. The transmission spectra of the 75 µm and 100 µm long interferometers used in this study are highlighted in bold. Similar spectral data was obtained from sensors of identical length. The data has not been processed with normalization or smoothing algorithms. (b) The median optical transmission value obtained from the spectra in Figure S2(a) plotted against their corresponding sensor lengths. The data shows a clear scaling in the transmitted optical power associated primarily to the plasmonic excitation and plasmonic-length dependent propagation.

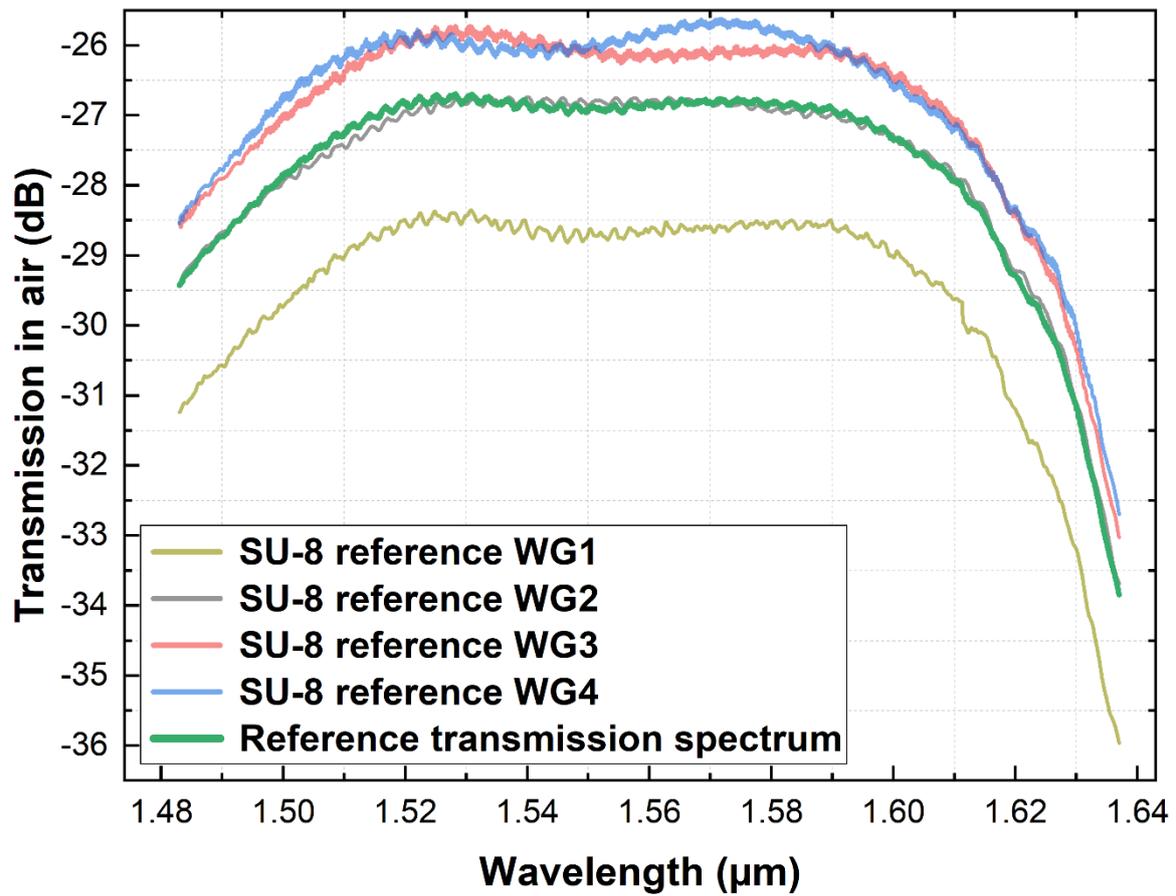

S3: The optical transmission spectrum of SU-8 WGs in air. These reference SU-8 WGs did not contain plasmonic stripes. The average spectrum (shown in bold) obtained from the reference WGs was utilized to calculate the sensor insertion loss reported in Figure 3e and Figure 3f.

## Simulations

The design analysis of the proposed plasmo-photonic sensor was carried out in two steps considering water as the ambient environment. The first step concerned the optimization process of the extinction ration (ER), and the second step involved the simulation of the complete sensor to obtain the expected transfer function using the INTERCONNECT photonic integrated circuit simulator of ANSYS Lumerical. To maximize the ER value at the sensor output, we deployed the definition of the ER according to equation (2) provided in the main text and the bimodal interferometric transfer function according to equation (5) as follows:

$$E_{out} = t_{01} e^{-j\beta_1 L} t_{10} + t_{02} e^{-j\beta_2 L} t_{20} \qquad (5)$$

where $E_{out}$ is the electric field at the output of the bimodal sensor, $t_{01}$, $t_{02}$ symbolize the electric field coupling coefficients from the photonic mode (0) to the top (1) and bottom (2) plasmonic modes respectively, $t_{10}$, $t_{20}$ symbolize the electric field coupling coefficients from the top and bottom plasmonic modes to the photonic mode, $L$ represents the plasmonic stripe length, and $\beta_1$, $\beta_2$ are the corresponding propagation constants of the top and bottom plasmonic modes, with $\beta_i = (2\pi/\lambda_0)(n_i - jk_i)$ and $n_i$, $k_i$ the real and imaginary parts of the complex effective index. Assuming no mode conversion at the plasmonic part, the coupling coefficients from the photonic to the plasmonic modes can be considered equal to the coupling coefficients from the plasmonic modes to the photonic mode ($t_{10} = t_{01}$ and $t_{20} = t_{02}$), and equation (5) can be simplified to:

$$E_{out} = |t_1|^2 e^{-j\beta_1 L} + |t_2|^2 e^{-j\beta_2 L} \qquad (6)$$

where $|t_1|^2$ and $|t_2|^2$ now symbolize the power coupling coefficients from the photonic mode to the two plasmonic modes and backwards. Considering that $T_{\max} = |E_{\max}|^2$ is obtained when the two interfering modes (top and bottom) are in-phase, i.e., $L\Delta n/\lambda_0$=integer, and that $T_{\min} = |E_{\min}|^2$ is obtained when the two interfering modes are off-phase i.e., $L\Delta n/\lambda_0$=half integer, then:

$$T_{max(min)} = \left|\left(|t_1|^2 \cdot e^{-a_1 L/2} \pm |t_2|^2 \cdot e^{-a_2 L/2}\right)\right|^2 \tag{7}$$

where $a_i = 4\pi k_i/\lambda_0$, $i$=1,2 represent the corresponding absorption coefficients and the sign +(-) is used for the max(min) values.

The propagation losses of the two plasmonic modes are different, so optimal interference and maximum ER can be attained by adjusting the power coupling coefficients between the photonic and plasmonic modes varying the thickness of the bottom SU-8 layer, in conjunction with selecting a suitable length for the plasmonic stripe. For this reason, a comprehensive 2D eigenmode analysis was conducted encompassing two key components: (i) the overlap integral analysis between the TM photonic mode and the two plasmonic modes that gives an estimation of the power fraction that can be coupled from the photonic to the plasmonic modes and back, and (ii) the evaluation of the propagation losses of the top and bottom plasmonic modes, with both being investigated in relation to the thickness of the bottom SU-8 layer. After that, the calculation of the ER by means of equations (1) and (7) was followed and the obtained results are presented in Figure *S4* for SU-8 thicknesses between 0.5 µm - 1.3 µm and for plasmonic lengths ranging from 10 µm - 200 µm. Figure S5 illustrates the ER as a function of the bottom layer thickness for the plasmonic lengths of 75 µm and 100 µm. As it can be seen, the maximum ER is obtained when the bottom SU-8 layer is 0.66 µm thick for both plasmonic lengths. The simulated ER of the 75 µm and 100 µm sensors fabricated on a 0.9 µm thick bottom SU-8 layer is 12.5 dB and 15 dB, respectively.

After the ER optimization, a full circuit level analysis was conducted using the INTERCONNECT photonic integrated circuit simulator of ANSYS Lumerical. The individual components of the entire sensor were modelled considering the broadband material properties and the WG dispersion. Two 'Waveguide' building blocks carrying the frequency-dependent properties (effective index, group index etc.) of the two plasmonic modes calculated in MODE WG simulator of Lumerical were used to model the sensing and reference branch of the bimodal interferometer. Two 'Y coupler' building blocks were used as a splitter at the input and as a combiner at the output of the structure. The insertion losses and the coupling coefficients of the 'Y coupler' building blocks were defined through overlap integral and 2D eigenmode analysis. Figure S6 depicts the spectral response in air obtained from these simulations for plasmonic stripe lengths of 75 µm and 100 µm.

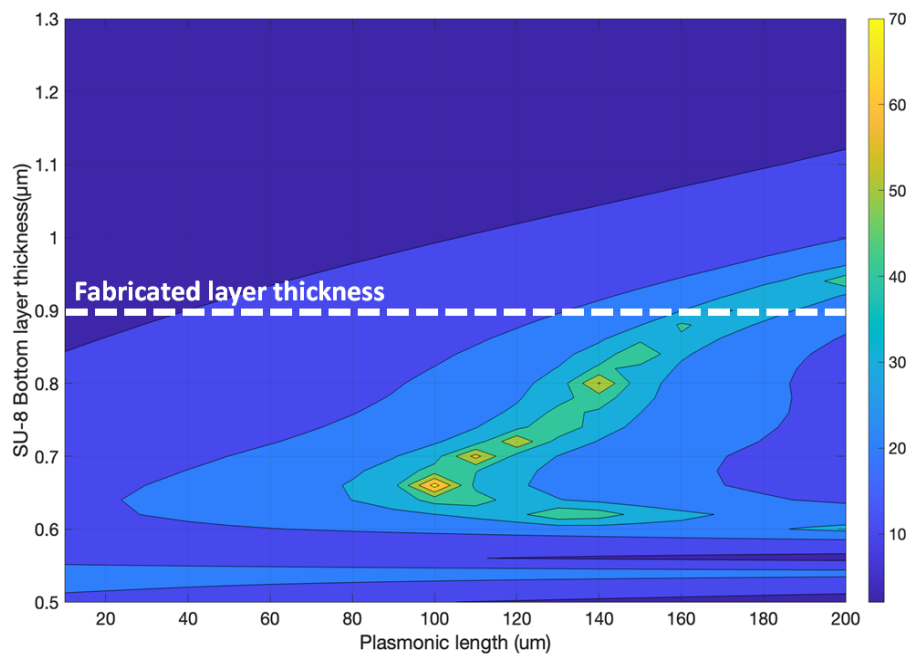

S4: Theoretical ER values for different thicknesses of the bottom SU-8 layer and for different plasmonic lengths.

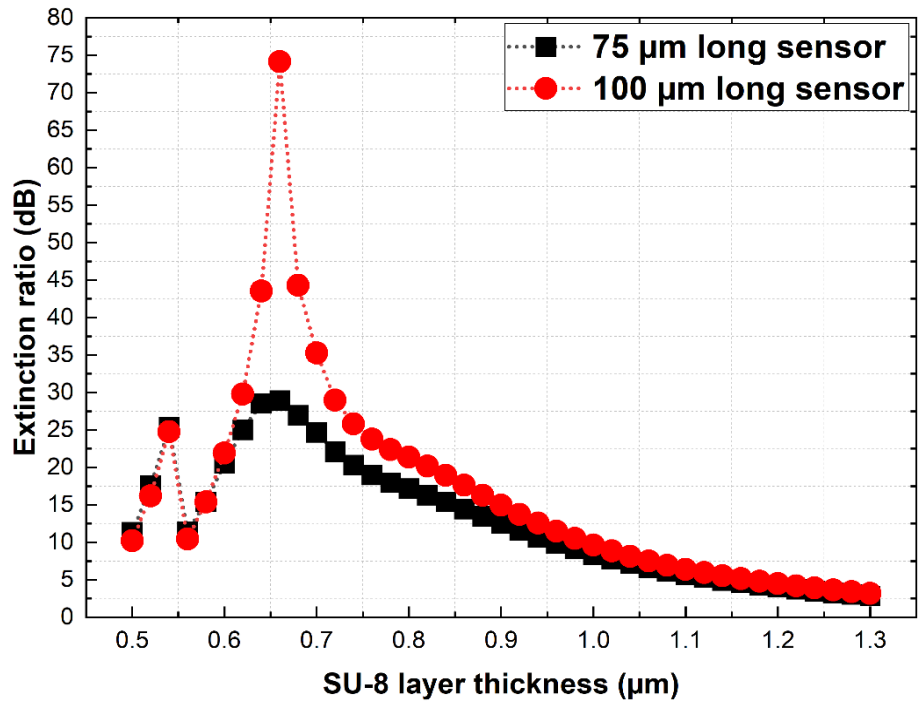

S5: Theoretical ER values as a function of the bottom SU-8 layer for 75 μm and 100 μm sensor lengths. The fabricated layer thickness of 0.9 μm corresponds to approximately 15 dB and 12 dB ER for the 75 μm and 100 μm sensors, respectively.

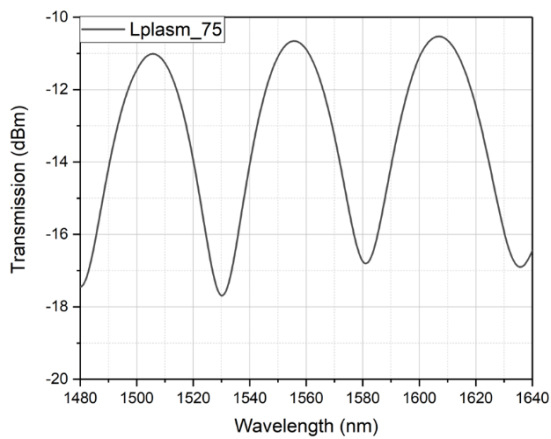 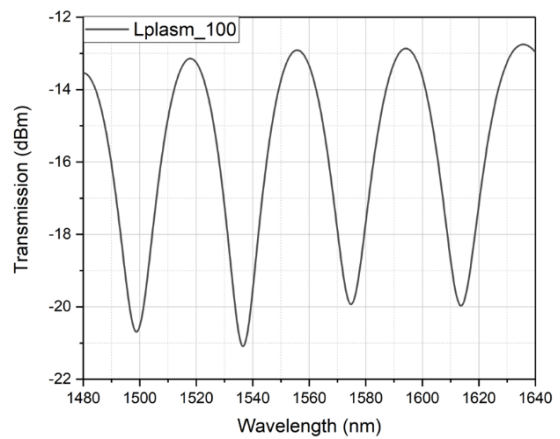

S6: Simulated transfer functions of the proposed bimodal sensor configurations of 75 μm (left) and 100 μm (right) plasmonic stripe lengths. The thickness of the bottom SU-8 layer was 0.9 μm.